# Synthesis and characterization of cotton fiber-based nanocellulose

T. Theivasanthi[1*], F.L. Anne Christma[2], Adeleke Joshua Toyin[3], Subash C.B. Gopinath[4,5], Ramanibai Ravichandran[6]

1. International Research Centre, Kalasalingam University, Krishnankoil 626126, Tamil Nadu, India
2. Department of Nanoscience and Nanotechnology, Karunya University, Tamil Nadu, India
3. Osun State University, Osogbo, Nigeria
4. School of Bioprocess Engineering, Universiti Malaysia, 02600 Arau, Perlis, Malaysia
5. Institute of Nano Electronic Engineering, Universiti Malaysia, 01000 Kangar, Perlis, Malaysia
6. Department of Zoology, University of Madras, Guindy Campus, Chennai, Tamil Nadu, India

*Corresponding author E-mail: ttheivasanthi@gmail.com   ORCID  : https://orcid.org/0000-0002-2280-9316

**Abstract:** Nanocellulose prepared from the natural material has a promising wide range of opportunities to obtain the superior material properties towards various end-products. In this research, commercially available natural cotton was treated with aqueous sodium hydroxide solution to eliminate the hemicellulose and lignin, then cellulose was collected. The collected cellulose was subjected to acid hydrolysis using sulphuric acid to obtain nanocellulose. The prepared nanocellulose was further characterized with the aid of Fourier transform infrared spectroscopy, X-ray diffraction and Scanning Electron Microscopy to elucidate the chemical structure, crystallinity and the morphology.

## 1. Introduction

For the past few decades, researchers have paid an attention to nanoscience and nanotechnology, because of its dynamic properties and applications [1–4]. Nanotechnology ("nanotech") deals the manipulation on an atom and molecular scale. It was subsequently reached a clear level with one dimension size ranged from 1 to 100 nanometers [5–9]. Nanotechnology and Nanoscience deal the ability to observe and to control the individual atom and molecule. Nanomaterials from cellulose play a major role in the nanotechnology field. Currently, researchers are finding different ways to deliberately create materials at the nano-level to be used their advantages, including lighter weight, higher strength, and good chemical reactivity than their large-scale counterparts and enhanced control of light spectrum.

Now-a-days, one of the hottest super materials is nanocellulose (NC) and widely been used. It has several applications, like pharmaceutical, food, electronics, due to its light, transparent and strong property. NC can be obtained from any natural cellulosic source materials, such as wood pulp, which consists of a tightly packed array of a needle-like structure called 'nanofibril'. The preparation strategy is systematic and obtained by top-down and bottom-up approaches. It is produced from the plant matter that has been reduced into small pieces and purified followed by homogenized to remove the non-homogenous compounds like lignin [10]. The creation of nanocellulose is completely in a neutral manner. Although NC offers diverse physical and chemical properties, even it makes the salt water to be drinkable. There is increasing interest in the use of cellulose nanofibre. Further, due to their higher aspect ratio, cellulose nanofibers have greater reinforcing capability than the currently used macro/microfibre feedstocks.

Cellulose is the major biodegradable polymer abundant on the earth. It is found in structural components of the cell wall of all green plants [11]. It is a polysaccharide compound and the homopolymer of glucose. A large number of glucose units combines to form a cellulose polymer molecule, which depends on their chain length and rate of polymerization [12]. The structure of cellulose is (β 1→ 4) linkage [13] as shown in Fig.1 a & b. It shows a very low conductivity and high resistivity of electricity. Based on the solubility nature, cellulose is classified into three categories. They are α, β and γ. Among them, α is insoluble, β is precipitate in nature and γ is completely soluble. Cellulose is found to be rich in natural cotton [14].

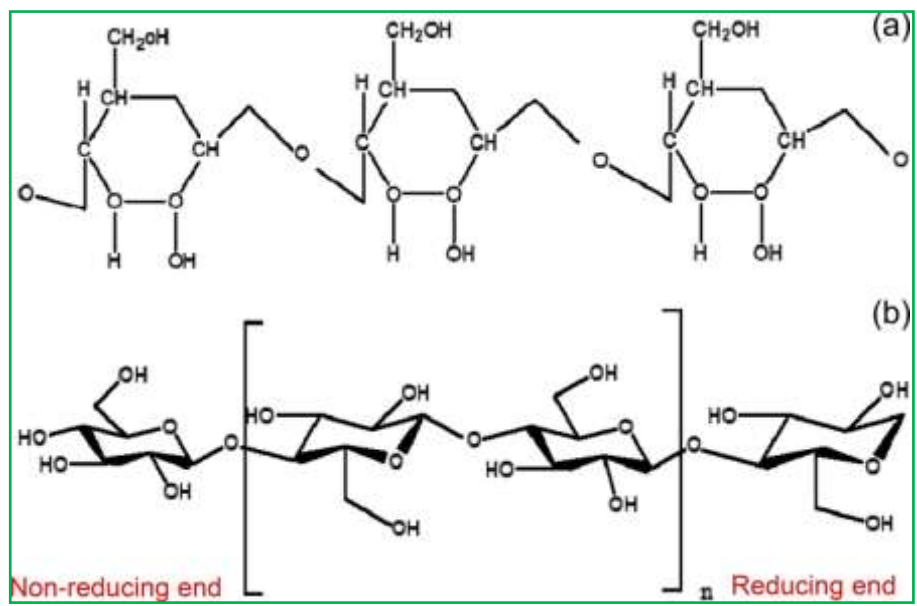

**Fig. 1.** Structures of cellulose. (a) Natural cellulose; (b) Nanocellulose

Cotton is the soft small ball-like fibrous material produced in the protective case. The cotton plant is the shrub and it contains large cellulose. Cotton producing places in the world are located in America, Asia, and Africa and the largest cotton production is from China [13]. The natural fibre used to produce NC is cotton, which has many advantages like wide availability, low-cost, renewable, abundant, strong, durable and biodegradable. Cotton swills in a high humid environment, water and in certain concentrated solutions [15–17].

Acid hydrolysis is the common method used for the preparation of NC [18]. Acid hydrolysis of cotton produces the hydrocellulose, which is not affected by cold weak acids. In acid hydrolysis, the most important considered point is cellulose acid ratio. The present research uses sulphuric acid-mediated synthesis and characterized the NC after acid hydrolysis method. Their physiochemical, structural and cellulose crystallinity index analyses were carried out by FTIR and XRD analysis.

## 2. Materials and Methods

All reagents procured were of analytical grade and used without additional purification steps. Commercially available Cotton was used as the source material. Sodium hydroxide pellets (NaOH) and sulphuric acid were received from Sigma-Aldrich. All experimental solutions were prepared using de-ionized (DI) water.

## 2.1. Synthesis of nanocellulose

The synthesis of NC follows a systematic procedure. Commercially available cotton was initially made into powder without impurities. The powdered sample was then treated with alkali solution NaOH solution. After this process cellulose was obtained. This sample was further subjected to acid hydrolysis. During the entire process of preparation, DI water was used as the solvent and stirred for ensuring uniform mixing. The pH was checked at every step and maintained as neutral. Fig. 2 exhibits the synthesis of nanocellulose.

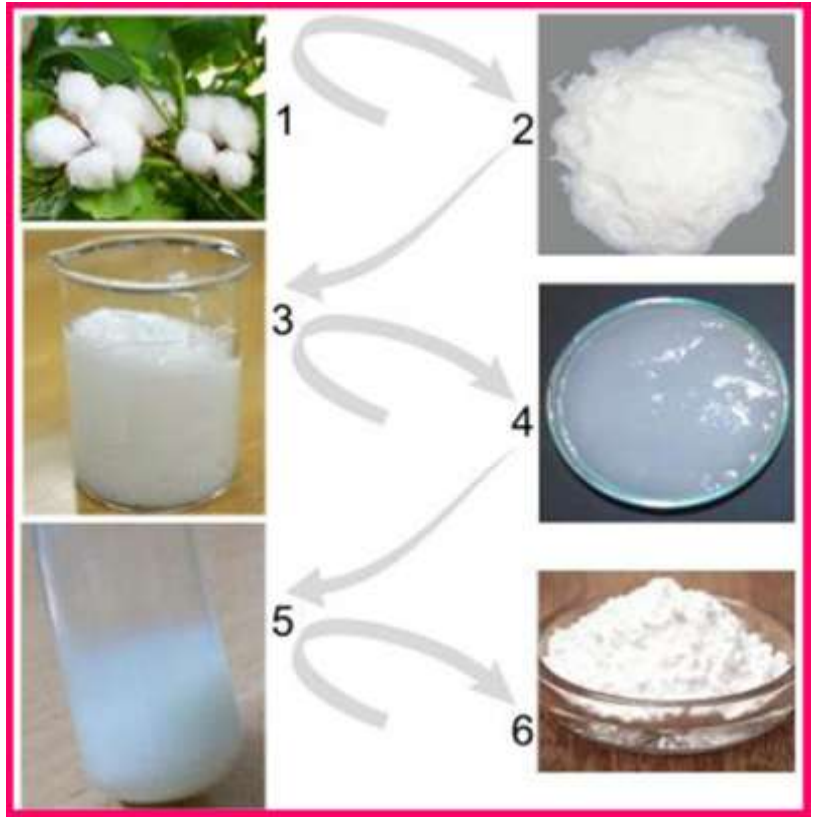

**Fig. 2.** Synthesis of Nanocellulose. (1) Raw cotton; (2) Cotton powder; (3) after Alkali treatment; (4) after Acid hydrolysis; (5) Nanocllulose; (6) Nanocellulose powder.

## 2.2. Alkali treatment on cotton

Commercially available Cotton was brought and ground in order to make a powder. 5% NaOH was added to the sample and was subjected to stirring constantly at room temperature for 4 h to get homogenous mixing. Then, it was washed and filtered several times with DI water adjusted to a neutral pH for the removal of lignin and hemicelluloses. The filtrate was dried at 80 °C for a day.

## 2.3. Acid hydrolysis

The alkali treated cotton was then added to the 10 ml of concentrated sulphuric acid and 20 ml of DI water, heated at 40 °C with constant stirring for an hour to get a well-mixed solution. It was then washed several times with the water adjusted to a neutral pH. The prepared suspended solution was centrifuged at 10,000 × g for 15 min. The obtained nanocellulose was dried at 80 °C in order to obtain a nanocrystalline powder.

### 2.4. Characterization of nanocellulose

The obtained cellulose was monitored by Fourier transform infrared spectroscopy (FTIR), Xray diffraction (XRD) and Scanning Electron Microscopy (SEM). FTIR (Spectrum 65, Perkin Elmer, USA) analysis was from 4000 to 500 cm$^{-1}$. The index crystallinity of the sample was measured by X-ray diffractometer (D8 advance ECO XRD system with SSD160 1D Detector) using Cu-Ka radiation wavelength of 1.5406–1.54439 Å. The samples were scanned in 2θ range of 15–70°. SEM observation (ZEISS DSM 940A) was at 20 kV.

## 3. Results and discussion

Nanocellulose (NC) is the polymer material that can be obtained cheaply from the plant sources. NC has a wide range of daily applications, including as foam, cloth, food, paper, and safety materials. In addition, in the industrial sector NC plays a pivotal role with medicine, hygiene and adsorbent products, emulsion, cosmetic and pharmaceutical. Recent developments in the nanotechnology brought enormous developments with NC-based materials and made significant achievements. In this study, nanocellulose from the natural fibre has been generated by treating with aqueous sodium hydroxide solution followed by the acid hydrolysis. NC was obtained by following the methods outlined above and subjected to FTIR, XRD and SEM analyses to characterize the chemical group, crystallinity and the surface morphology.

### 3.1. FTIR analysis

Infrared spectroscopy works based on the atoms' vibrations in a molecule to be tested. When the test molecule absorbs infrared radiation, the chemical bonds in it vibrate and able to stretch, contract or bend. Infrared spectra of the controlled sample displayed in the out of plane O-H stretching at 3342.64 cm$^{-1}$; C-O stretching at 1050–1120 cm$^{-1}$; The vibration of C=O carbonyl ring stretching occurred at 1728.22 cm$^{-1}$. C-O and C-H bonds are in the polysaccharide aromatic rings, whereas alkali treatment minimizes the hydrogen bonding by eliminating hydroxyl groups by a reaction with NaOH [19]. C-H stretching occurred at 2900–2970 cm$^{-1}$. The C-O vibration stretching resulted from the uronic and acetyl ester groups; from hemicellulose, pectin or the ester of carboxylic group of p-coumaric and ferulic acids of hemicellulose or lignin. C-O-C pyranose ring was attributed to the structure of cellulose its stretching vibration is at 1058.9 cm$^{-1}$ [20]. In cellulosic samples, the spectral ratio of 1431.18 cm$^{-1}$ provides the evidence of containing cellulose (Fig. 3).

### 3.2. XRD analysis

Fig.4 shows the X-ray diffraction pattern of NC. From this analysis, the high intensity peak was noticed at 2θ = 22.8° (arising due to diffraction from (200) plane), with extra double peak signals at = 14.9° and 16.5° due to reflections from (110) and (110) diffraction, respectively (Fig. 4). The dried NC was observed with nearly 91.2% crystalline and the XRD traces displayed a clear retention with the cellulose crystallites and also enhancing peak intensity at 2θ = 14°, 16°, 22°, 34° [21]. The more value of the given index shows that test material has high crystalline with a structure of proper order. Due to the treatments with chemical, may disturb the order in the structure and in the case of NC reached a higher value due to the removal of the amorphous state of cellulose.

Crystallinity Index or CrI was calculated by using the Eq (1),

$$CrI = \frac{I_{(002)} - I_{(am)}}{I_{(002)}} \times 100 \quad \ldots\ldots\ldots\ldots\ldots\ldots\ldots\ldots (1)$$

Where, $I_{(002)}$ is peak intensity of crystalline part and $I_{(am)}$ is counter reading at a peak intensity of the amorphous material.

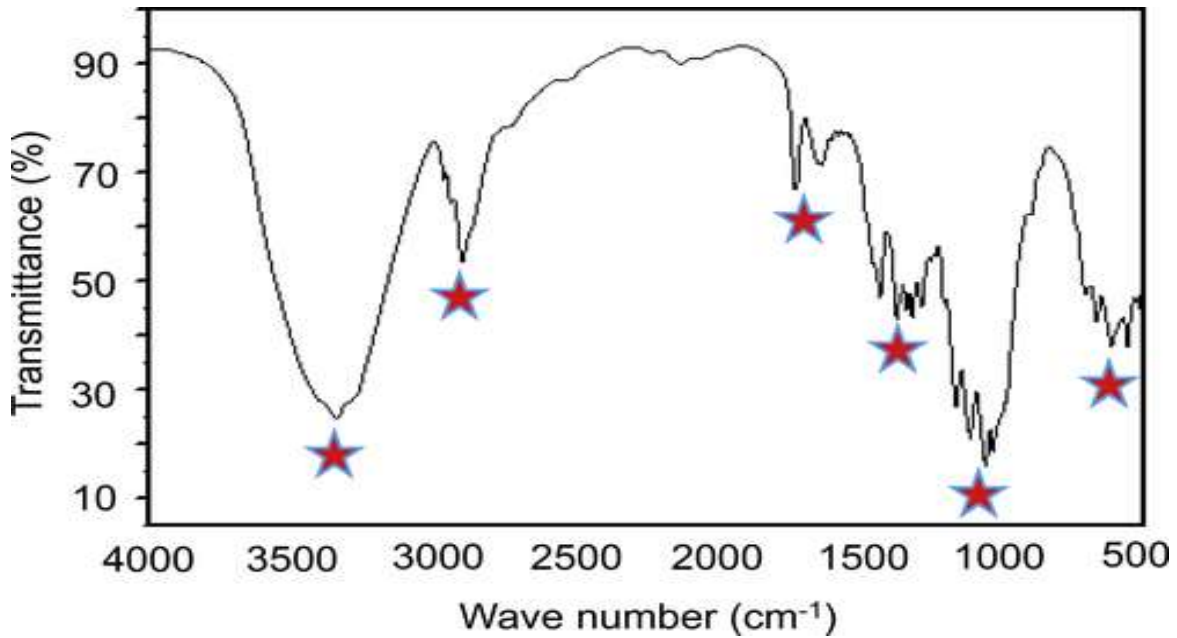

**Fig. 3.** FTIR spectra of nanocellulose. Analysis was from 4000 to 500 cm$^{-1}$.

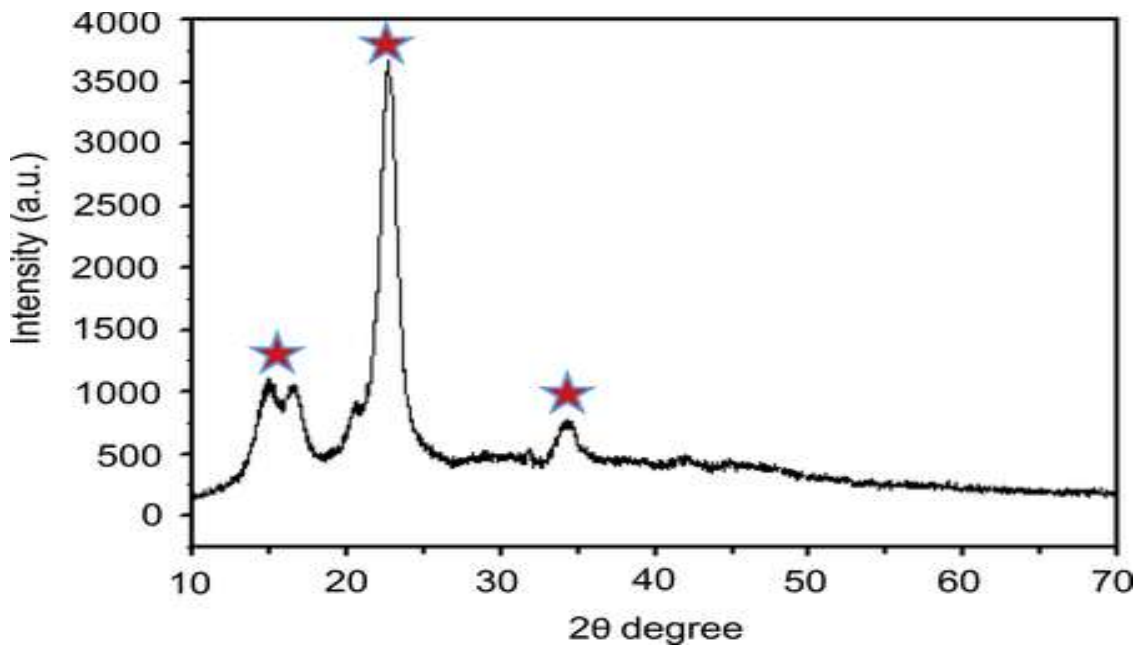

**Fig. 4.** XRD pattern of NC. Cu-Kα radiation wavelength of 1.5406–1.54439 A° was used. Samples were scanned in 2θ range of 15–70°.

### 3.3. SEM observation

The surface morphology of the cotton samples was monitored by a Zeiss DSM 940 A under accelerated electrons with 15 kV of energy using SEM. The shape of this fibre indicates an increase in its specific area, favours chemical reaction such as acid hydrolysis. The rod shape increases the area of the surface and creates the fibre to be more reactive (Fig. 5).

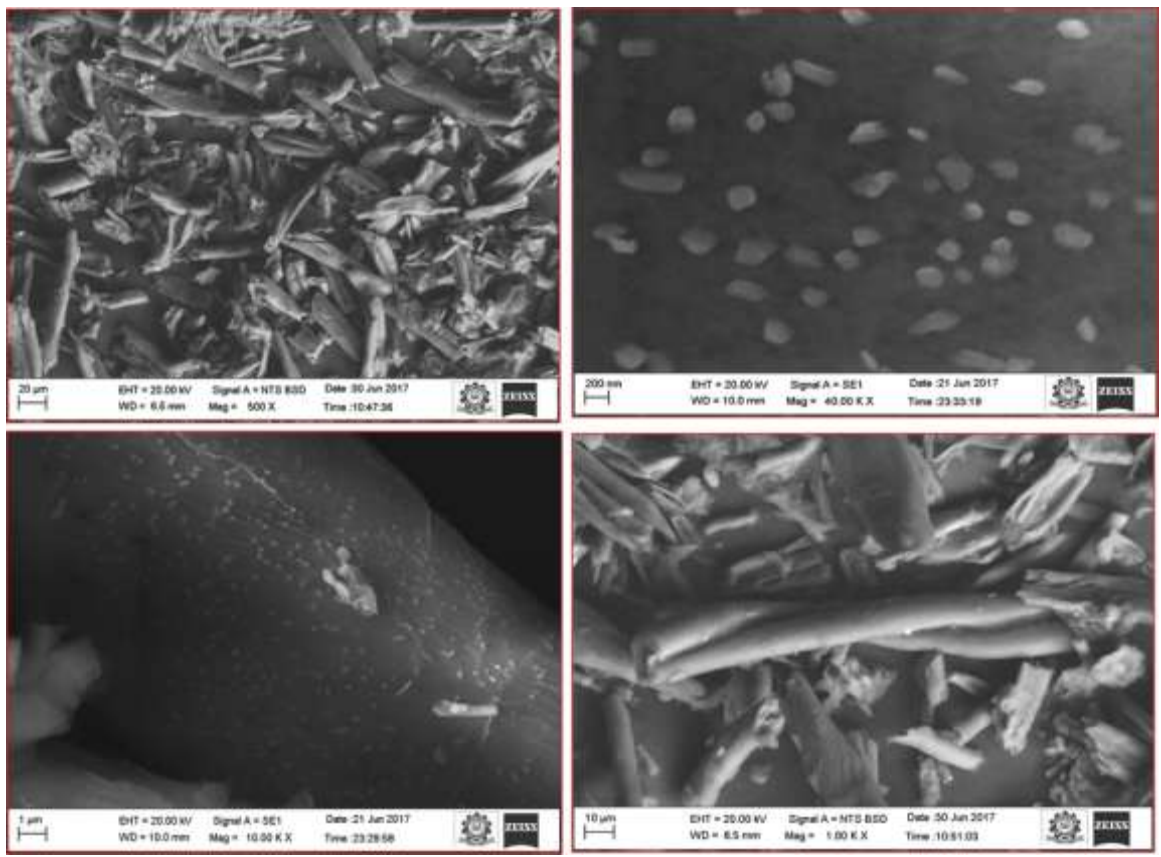

**Fig. 5.** SEM images of the obtained nanocellulose. Observed under different magnifications at 20 kV.

### 4. Conclusions

NC was successfully extracted from the natural cotton and the characterization results with Fourier transform infrared spectroscopy, Scanning Electron Microscopy and X-ray diffraction analyses on the cellulose material show a good level of purity and high crystallinities. The crystallinity index for the obtained NC is 91.7%. Utilization of sodium hydroxide concentration promoted the preservation of the structure of cellulose and favoured the hydrolysis of the components with the amorphous state and the cost-efficient products from cotton. The result proves that cotton can be used to produce NC which in turn can be utilized for downstream applications.

**Acknowledgment**

Authors are grateful to International Research Centre, Kalasalingam University, for providing analytical and research facilities.